\documentclass[12pt]{article}

\pdfoutput=1

\usepackage{graphicx}
\usepackage{epsfig}
\usepackage{amsfonts}
\usepackage{amssymb}
\usepackage{amsmath}
\usepackage{dsfont}
\usepackage{bbm}
\usepackage{multirow}
\usepackage[vcentermath]{youngtab}
\usepackage{latexsym}
\usepackage{verbatim}
\usepackage{mcite}
\usepackage{cite}
\usepackage{units}

\def\beq{\begin{equation}}
\def\eeq{\end{equation}}
\def\beqa{\begin{eqnarray}}
\def\eeqa{\end{eqnarray}}

\def\bfone{\relax{\rm 1\kern-.35em 1}}

\newcommand{\be}{\begin{equation}}
\newcommand{\ee}{\end{equation}}
\newcommand{\ben}{\begin{displaymath}}
\newcommand{\een}{\end{displaymath}}
\newcommand{\bea}{\begin{eqnarray}}
\newcommand{\eea}{\end{eqnarray}}

\newcommand{\bean}{\begin{eqnarray*}}
\newcommand{\eean}{\end{eqnarray*}}

\begin{document}
\pagestyle{plain}

%----------------------------------------------------------------------%
%  numbering sections, equations, footnotes, etc...
%----------------------------------------------------------------------%

\makeatletter
\@addtoreset{equation}{section}
\makeatother
\renewcommand{\thesection}{\arabic{section}}
\renewcommand{\theequation}{\thesection.\arabic{equation}}
\renewcommand{\thefootnote}{\arabic{footnote}}

%----------------------------------------------------------------------%
%  Resetting of counters
%----------------------------------------------------------------------%

\setcounter{page}{1}
\setcounter{footnote}{0}

\begin{center}

\vskip 0cm

{\LARGE \bf Vacua Analysis in Extended Supersymmetry\\[2mm] Compactifications} 
\vskip 25pt 

{\bf Giuseppe Dibitetto\footnote{Corresponding author $\,$ E-mail:~\textsf{g.dibitetto@rug.nl}, Phone: +31$\,$ (0)50$\,$363 9043}, Adolfo Guarino and  Diederik Roest}

\vskip 0.3cm

{\em Centre for Theoretical Physics,\\
University of Groningen, \\
Nijenborgh 4, 9747 AG Groningen, The Netherlands}

\vskip 0.8cm

{\bf ABSTRACT}\\[3ex]
\end{center}

%\begin{minipage}{13cm}
%\small

We analyse geometric type IIA flux compactifications leading to $\mathcal{N}=4$ gauged supergravities in four dimensions. The complete landscape of isotropic vacua is presented, which turns out to belong to a unique theory. The solutions admit an uplift to maximal supergravity due to the vanishing of the flux-induced tadpoles for all the supersymmetry-breaking branes. Such an uplift is sketched out and the full $\mathcal{N}=8$ mass spectra are discussed. We find the interesting presence of a non-supersymmetric and nevertheless stable minimum.
%\end{minipage}

\section{Introduction}

Half-maximal and maximal gauged supergravities in four dimensions are the low-energy effective theories arising from flux compactifications in string theory, provided that only internal manifolds and extended objects are included which are compatible with such amounts of supersymmetry. In the last decade the embedding tensor formalism has been used extensively in the context of (half-)maximal supergravity in order to describe all the deformations of the free theory in a duality-covariant way. Nevertheless, it has already been pointed out in the literature that not all the gaugings of supergravity have a higher-dimensional origin in terms of a geometric flux compactification in string theory.  This indicates that gaugings coming from geometric flux compactifications are not a closed set under general duality transformations and this is the origin of non-geometric fluxes \cite{hep-th/0508133}. Gaugings associated with such fluxes might yet have a higher-dimensional description in terms of a double field theory (DFT) \cite{Aldazabal:2011nj, Geissbuhler:2011mx}, in which duality covariance becomes the fundamental principle to start with, independently of the compactification procedure. Because of this interpretation, gauged supergravities with extended supersymmetry seem to be suitable frameworks for investigating the nature of non-geometric fluxes, as summarised in table~\ref{duality}.

\begin{table}[h]
\renewcommand{\arraystretch}{1.25}
\begin{center}
\scalebox{0.85}[0.85]{
\begin{tabular}{|c|c|c|}
\hline
SUSY & $G$ & stringy interpretation \\
\hline \hline
$\mathcal{N}=4$ &   SL($2$)$\,\times\,$SO($6,6$) & S- and T-duality\\
\hline
$\mathcal{N}=8$ &   E$_{7(7)}$ & U-duality\\
\hline
\end{tabular}
}
\end{center}
\caption{Half-maximal and maximal gauged supegravities in four dimensions seem to be suitable playgrounds to understand how to restore duality covariance in flux compactifications.}

\label{duality}

\end{table}

\vspace{-2mm}

After introducing the SO($3$) truncation of half-maximal supergravity as the effective theory arising from specific type IIA orientifold reductions with background fluxes, we analyse the landscape of isotropic vacua in geometric backgrounds, i.e. including metric and gauge fluxes. We find a set of anti-de Sitter (AdS) critical points admitting an uplift to $\mathcal{N}=8$, one of which is remarkably stable without preserving any supersymmetry. More details on this work can be found in refs~\cite{Dibitetto:2011gm, arXiv:1104.3587, DGRmax}.
The goal of finding de Sitter (dS) vacua motivates the analysis of non-geometric flux compactifications as possible future extensions.

\section{The geometric type IIA with O6/D6 setup}

Upon the SO($3$) truncation, half-maximal supergravity in four dimensions reduces to a three-field $STU$ model in the following way
\begin{center}
SL($2$)$\,\times\,$SO($6,6$)$\qquad\longrightarrow\qquad$
SL($2$)$\,\times\,$SO($2,2$) $\,\,\,=\,\,\,$
SL($2$)$_S\,\times\,$SL($2$)$_T\,\times\,$SL($2$)$_U$ \,\,.
\end{center}
The truncation allows for forty SO($3$)-singlet embedding tensor components that can be written in terms of an SL($2$) $\times$ SO($2,2$) tensor $\Lambda_{\alpha ABC} = \Lambda_{\alpha (ABC)}$, with $\alpha=\pm$ and $A=1,...,4$ being SL($2$) and SO($2,2$) fundamental indices, respectively. As shown in ref.~\cite{Dibitetto:2011gm}, these forty embedding tensor components correspond exactly with the set of generalised fluxes in type II orientifold reductions on a $\mathbb{Z}_2\times\mathbb{Z}_2$ isotropic orbifold. As a consequence of the truncation, the theory preserves only $\mathcal{N}=1$ supersymmetry out of the original $\mathcal{N}=4$. Therefore, the corresponding scalar potential can be written in terms of a (logarithmic) K\"ahler potential $K$ and a holomorphic superpotential $W$ by using the standard $\mathcal{N}=1$ expression
\begin{equation}\label{VN1}
V^{(\text{SO}(3))} = e^K \left( - 3|W|^2 + |D\, W|^2 \right)\,.
\end{equation}
At the effective level, fluxes appear as arbitrary superpotential couplings up to linear in $S$ and up to cubic in $T$ and $U$. The scalar potential computed from \eqref{VN1} turns out to coincide with the scalar potential given in ref.~\cite{Schon:2006kz} up to terms projected out by a set of $\mathcal{N}=4$ quadratic constraints on the embedding tensor $\Lambda$ \cite{Dibitetto:2011gm}.

\begin{table}[h]
\renewcommand{\arraystretch}{1.25}
\begin{center}
\scalebox{0.85}[0.85]{
\begin{tabular}{ | c || c | c | c | c | c |}
\hline
$W$ couplings  & Type IIA fluxes & Embedding tensor components\\
\hline \hline
$a_0$ & $F_6$ & $-\Lambda_{+333}$\\
\hline
$a_1$ & $F_4$ & $\Lambda_{+334}$\\
\hline
$a_2$ & $F_2$ & $-\Lambda_{+344}$\\
\hline
$a_3$ & $F_0$ & $\Lambda_{+444}$\\
\hline
$b_0$ & $H_3$ & $-\Lambda_{-333}$\\
\hline
$b_1$ & $\omega$ & $\Lambda_{-334}$\\
\hline
$c_0$ & $H_3$ & $\Lambda_{+233}$\\
\hline
$c_1\,,\,\tilde{c}_1$ & $\omega$ & $\Lambda_{+234}\,,\,\Lambda_{+133}$\\
\hline
\end{tabular}
}
\end{center}
\caption{The set of SO($3$)-invariant embedding tensor components of $\Lambda$ admitting a higher-dimensional origin as type IIA fluxes: a metric flux $\omega$ together with R-R $F_{0,2,4,6}$ and NS-NS $H_{3}$ gauge fluxes.}

\label{fluxes}

\end{table}
\vspace{-2mm}

Restricting to the components of $\Lambda$ that can be interpreted as metric or gauge fluxes in a type IIA realisation of the model, we are left with the following superpotential
\begin{equation} 
W_{\text{IIA}}=a_0 - 3 \, a_1 \, U + 3 \, a_2 \, U^2 - a_3 \,
U^3 - b_0\,S + 3 \, b_1 \, S\,U + 3 \, c_0 \, T + (6 \, c_{1} - 3\,
\tilde{c}_{1}) \, T\,U \, , 
\end{equation}
consisting of nine flux-induced couplings (see table ~\ref{fluxes}). These fluxes are demanded to satisfy the set of $\mathcal{N}=4$ quadratic constraints
\begin{equation} \label{QC}
\begin{array}{ccr}
c_{1}\,(c_{1}-\tilde{c}_{1})=0 \hspace{7mm} , \hspace{7mm}
b_{1}\,(c_{1}-\tilde{c}_{1})=0 & & (\omega^2\,=\,0)\,\,\,,\\
- a_{3} \, c_{0} - a_{2} \, (2\, c_{1}-\tilde{c}_{1})=0 & &
({N_6}^{\bot}\,=\,0)\,\,\, ,
\end{array}
\end{equation} 
where the first two constraints are related to the nilpotency of the twisted exterior derivative operator, whereas the third one imposes the absence of D6-branes wrapping the directions orthogonal to the O6-planes, which would break supersymmetry explicitly down to $\mathcal{N}=1$. In contrast, D6-branes parallel to the O6-planes are compatible with $\mathcal{N}=4$ supersymmetry, hence being allowed. Their corresponding flux-induced tadpole reads
\begin{equation} \label{tadpole}
{N_{6}}^{||}= 3 \, a_{2} \, b_{1} - a_{3} \, b_{0}\,=\,\frac{N_{O6} }{2} - N_{D6}\, . 
\end{equation} 

After observing that the set of fluxes given in table \ref{fluxes} is closed under non-compact duality transformations (i.e. real shifts and rescalings of $S$, $T$ and $U$), we can restrict the search for critical points of the scalar potential to the point $S_{0}=T_{0}=U_{0}=i$ without losing generality. The field equations become then quadratic conditions in the fluxes that have to be satisfied together with the quadratic constraints in \eqref{QC}. These equations generate a quadratic ideal which we decompose in terms of prime ideals by using the Gianni-Trager-Zacharias (GTZ) decomposition \cite{GTZ} with the help of \textsc{Singular} \cite{DGPS}. By solving them, the complete set of critical points of the scalar potential is presented in table \ref{table:N=4_vacua}. They turn out to be (modulo the discrete $\mathbb{Z}_2$ symmetry introduced in the caption) different AdS critical points of a unique theory with an underlying gauging given by the gauge group $G_{0}=$ISO($3$)$\,\ltimes\,$U($1$)$^6$.
\begin{table}[h]
\renewcommand{\arraystretch}{2.00}
\begin{center}
\scalebox{0.77}[0.77]{
\begin{tabular}{ | c || c | c |c | c | c | c |c | c |}
\hline
\textrm{\textsc{id}} & $a_{0}$ & $a_{1}$ & $a_{2}$ & $a_{3}$ & $b_{0}$ & $b_{1}$ & $c_{0}$ & $c_{1}=\tilde{c}_{1}$ \\[1mm]
\hline \hline
$1_{s}$ & $s \,  \dfrac{3 \,\sqrt{10}}{2}\, \lambda $ & $ \,\dfrac{\sqrt{6}}{2} \, \lambda$ & $ - s \,\dfrac{\sqrt{10}}{6} \, \lambda$ & $ \, \dfrac{5\,\sqrt{6}}{6} \, \lambda$ & $- \,s \, \dfrac{\sqrt{6}}{3} \, \lambda$ & $\dfrac{\sqrt{10}}{3}\,\lambda$ & $s \, \dfrac{\sqrt{6}}{3}\,\lambda$ & $\sqrt{10} \, \lambda$   \\[1mm]
\hline \hline
$2_{s}$ & $s \,\dfrac{16 \, \sqrt{10}}{9} \,\lambda$ & $0$ & $0$ & $ \, \dfrac{16 \, \sqrt{2}}{9} \, \lambda$ & $0$ & $\dfrac{16 \, \sqrt{10}}{45} \, \lambda$ & $0$ & $\dfrac{16 \, \sqrt{10}}{15} \, \lambda$   \\[1mm]
\hline
$3_{s}$ & $s \,\dfrac{4\,\sqrt{10}}{5}\,\lambda$ & $- \, \dfrac{4\,\sqrt{30}}{15}\,\lambda$ & $s \, \dfrac{4\,\sqrt{10}}{15}\,\lambda$ & $s \,\dfrac{4\,\sqrt{30}}{15}\,\lambda$ & $\, s \,\dfrac{4\,\sqrt{30}}{15}\,\lambda$ & $\dfrac{4\,\sqrt{10}}{15}\,\lambda$ & $- \, s \,\dfrac{4\,\sqrt{30}}{15}\,\lambda$ & $\dfrac{4\,\sqrt{10}}{5}\,\lambda$   \\[1mm]
\hline
$4_{s}$ & $s \,\dfrac{16 \, \sqrt{10}}{9} \,\lambda$ & $0$ & $0$ & $\,\dfrac{16 \, \sqrt{2}}{9} \,\lambda$ & $0$ & $\dfrac{16 \, \sqrt{2}}{9} \,\lambda$ & $0$ & $\dfrac{16 \, \sqrt{2}}{9} \,\lambda$  \\
\hline
\end{tabular}
}
\end{center}
\caption{The set of critical points of the scalar potential for geometric type IIA isotropic flux compactifications. The solutions labelled with $1_s$ turn out to preserve $\mathcal{N}=1$ supersymmetry for $s=+1$ and to be non-supersymmetric for $s=-1$, whereas all the others are non-supersymmetric. It is worth noticing that $s=\pm\,1$ appears as an accidental $\mathbb{Z}_2$ symmetry which relates solutions having exactly the same energy and the same mass spectrum. The parameter $\lambda$ is a global scaling parameter such that $V \propto \lambda^{2}$.} 

\label{table:N=4_vacua}

\end{table} 
\vspace{-2mm}

Regarding stability at the critical points, we computed in ref.~\cite{Dibitetto:2011gm} the full mass matrix for the $38$ physical scalars in $\mathcal{N}=4$ making use of the results in ref.~\cite{Borghese:2010ei}. At this point, we can say two things: firstly, the solutions $1_s$ are fully stable because of (fake-) supersymmetry, and secondly the solutions $2_s$ are already unstable because of the presence of a tachyon whose mass is below the Breitenlohner-Freedman (BF) bound. However, this is not enough since all the critical points turn out to be compatible with the total absence of sources, i.e., the flux-induced tadpole \eqref{tadpole} does accidentally vanish at these points. As a result, they admit an uplift to $\mathcal{N}=8$ and hence one should analyse the full mass matrix for the $70$ scalars spanning the coset E$_{7(7)}$/SU($8$) in order to make any final statement about stability. We worked out the uplifting by embedding the $R$-symmetry group of half-maximal theory, i.e. U$(4)$ $=$ U$(1)$ $\times$ SU$(4)$, into that of the maximal, i.e. SU$(8)$, and relating the fermionic shifts given in ref.~\cite{Schon:2006kz} to those ones in ref.~\cite{de_Wit} according to the decomposition
%
%\vspace{-1mm}
%
\begin{equation}
\begin{small}
\textrm{E}_{7(7)}\,\begin{array}{c}\nearrow\\
\searrow\end{array}
\begin{array}{c}
\textrm{SL}(2)\times\textrm{SO}(6,6)\\
\phantom{blabla} \\
\phantom{blabla} \\
\textrm{SU}(8)
\end{array}
\begin{array}{c}
\searrow\\
\nearrow
\end{array}\qquad
\underbrace{\textrm{U}(1)\,\times\,\textrm{SU}(4)_{m}}_{R\textrm{-symmetry
of the }\mathcal{N}=4 \textrm{ theory}}\,\times\,\textrm{SU}(4)_{a}\,\,\, .
\label{Rsymmetry}
\end{small}
\end{equation}
\vspace{-2mm}

\noindent After this uplifting, the whole set of critical points are found to satisfy the equations of motion and the quadratic constraints of the maximal theory. Subsequently we computed the mass matrix for the scalars in the $\mathcal{N}=8$ theory using the results in ref.~\cite{LeDiffon:2011wt}. The results are summarised in table~\ref{table:stability}. We find that the solutions $3_{s}$ are non-supersymmetric and nevertheless stable. Up to our knowledge, this is the second example in the literature (after the one in ref.~\cite{arXiv:1010.4910}) of such a solution in maximal supergravity; as opposed to the first example, though, this solution is completely tachyon-free rather than presenting tachyons although still above the BF bound. A final remark is that now, what used to be an accidental $\mathbb{Z}_2$ symmetry in the solutions has a proper interpretation within the $\mathcal{N}=8$ theory, i.e. it interchanges SU($4$)$_m$ and SU($4$)$_a$ in eq.~\eqref{Rsymmetry}.  
\begin{table}[h]
\renewcommand{\arraystretch}{1.80}
\begin{center}
\scalebox{0.77}[0.77]{
\begin{tabular}{ | c || c | c |c |c | }
\hline
\textrm{\textsc{id}} & $V_0$ & ${m^{2}}_{(\mathcal{N}=4)}$ & ${m^{2}}_{(\mathcal{N}=8)}$  & \textrm{Stability}\\[1mm]
\hline \hline
$1_{s}$ &   $-\lambda^{2}$  &   $-\dfrac{2}{3}$   &   $-\dfrac{2}{3}$  & \textrm{\textsc{stable}} \\[1mm]
\hline \hline
$2_{s}$ &   $-\dfrac{32}{27}\,\lambda^{2}$   &  $ -\dfrac{4}{5}$   &    $-\dfrac{4}{5}$  & \textrm{\textsc{unstable}} \\[1mm]
\hline
$3_{s}$ &   $-\dfrac{8}{15}\,\lambda^{2}$   &   $0$   &  $ 0$  & \textrm{\textsc{stable}} \\[1mm]
\hline
$4_{s}$ &   $-\dfrac{32}{27}\,\lambda^{2}$    &  $ 0$   &  $ -\dfrac{4}{3}$  & \textrm{\textsc{unstable}} \\
\hline
\end{tabular}
}
\end{center}
\caption{The values of the energy and the normalised mass for the lightest scalar at the set of critical points. One observes that, when lifting from  $\mathcal{N}=4$ to $\mathcal{N}=8$, the solutions $4_s$ become unstable because of the appearance of an unstable tachyonic direction within the new scalar modes. We remind the reader that in four dimensions the BF bound is given by $m^{2}_{ \textrm{BF}}=-3/4$ in units of the scalar potential.} 

\label{table:stability}

\end{table}

\section{Conclusions}

The study of isotropic type IIA orientifolds including geometric fluxes and preserving half-maximal supersymmetry reveals the presence of only AdS vacua in the landscape. The solutions turn out to be all liftable to maximal gauged supergravity where they appear as four different critical points of a unique theory: one is supersymmetric and stable, another one is non-supersymmetry and nevertheless stable and the remaining two are non-supersymmetry and unstable. The natural extension of this work will be to study the effect of non-geometric fluxes in this setup in order to get a richer landscape, maybe even containing dS solutions.

\section*{Acknowledgements}
  We are grateful to Andrea Borghese for stimulating discussions. The work of the authors is supported by a VIDI grant from the Netherlands Organisation for Scientific Research (NWO). Furthermore, G.D.~would like to thank the organisers of the ``XVII European Workshop
on String Theory", held in September 2011 in Padua, for a very stimulating experience.

\end{document}